\documentclass[onecolumn,pre]{revtex4}
\usepackage{epsfig}% Include figure files
\usepackage{floatflt}
\usepackage{graphicx}
\usepackage{amsmath,amssymb,mathrsfs}
\usepackage{setspace}

\usepackage[T1]{fontenc}
\usepackage[latin1]{inputenc}
\usepackage[cyr]{aeguill}
\usepackage[francais]{babel}

\begin{document}

\title{Nonlinear dielectric susceptibilities in supercooled liquids: a toy model.}
\author{F. Ladieu}
\email{francois.ladieu@cea.fr}
\author{C. Brun}
\author{D. L'H\^ote}
\affiliation{SPEC/SPHYNX (CNRS URA 2464), DSM/IRAMIS CEA Saclay, Bat.772, F-91191 Gif-sur-Yvette  France}
\date{\today}
\begin{abstract}
The dielectric response of supercooled liquids is phenomenologically modeled by a set of Asymmetric Double Wells (ADW), where each ADW contains 
a dynamical heterogeneity of $N_{corr}$ molecules. We find that the linear macroscopic susceptibility $\chi_1$ does not depend on 
$N_{corr}$ contrary to all higher order susceptibilities $\chi_{2k+1}$. We show that $\chi_{2k+1}$ is proportional to the $k^{th}$ moment 
of $N_{corr}$, which could pave the way for new experiments on glass transition. In particular, as predicted by Bouchaud 
and Biroli on general grounds 
[Phys. Rev. B, {\bf 72}, 064204 (2005)], we find that $\chi_3$ is proportional 
to the average value of $N_{corr}$. We fully calculate $\chi_3$ and, with plausible values of few parameters our model  
accounts for the salient features of the experimental behavior of $\chi_3$ of supercooled glycerol.
\end{abstract}

\maketitle
 
Upon fast enough cooling, most liquids do not cristallize but enter into a supercooled liquid state \cite{Deb01,Don01,Ang88,Edi96,Ada65}, 
where the viscosity $\eta$ dramatically increases 
with lowering the temperature $T$. Below the glass transition temperature $T_g$, $\eta$ is so high that the system is in practice a solid -the glass-,
 yet, no structural difference between the glass and the liquid state has ever been detected \cite{Deb01}. Over the past fifteen years, a major
  breakthrough was the discovery of Dynamical Heterogeneities (D.H.) in supercooled liquids \cite{Hur95,Sch96,Tra98,Edi00,Ric02,Ric06}; 
  i.e., relaxation happens through collective events gathering $N_{corr}$ molecules, and some groups are 
    relaxing much faster than others. As it is expected that an increase of $N_{corr}$ when lowering $T$ could increase dramatically $\eta$, an 
     significant effort was made for measuring the $T$-dependence of $N_{corr}$ \cite{Tra98,Ber05,Dal07}. 

It has been argued that the most direct way to draw accurately the $T$-dependence of $N_{corr}$ from experimental data is based on the a.c. 
nonlinear susceptibility $\chi_3$ \cite{Bou05,Cra10,Bru11}, where $\chi_3$ is the third order response of the fluid to a field with an angular frequency $\omega$. This field can be of any nature, e.g. electric as in \cite{Cra10,Bru11}. More precisely 
two nonlinear susceptibilities are related to $\chi_3$: $\chi_3^{(3)}$ and $\chi_{3}^{(1)}$, which correspond to the third
 order nonlinear response at the third harmonics (i.e., 
  at ``$3\omega$'') and at the first harmonics (i.e., at ``$1\omega$'') respectively. 
  Bouchaud and Biroli (BB) have shown \cite{Bou05,Bru11} that $\chi_3^{(3)}$ and $\chi_3^{(1)}$
should be related to the average value of $N_{corr}$ over the various D.H.'s existing at a given $T$ -noted $[N_{corr}(T)]_{av}$- by:

\begin{eqnarray}\label{eq1}
\chi_3^{(3)}(\omega,T) & \approx & \frac{\epsilon_0 (\Delta \chi_1)^2 a^3}{k_BT} [N_{corr}(T)]_{av} {\cal H}\left(\omega \tau_{\alpha}\right)\qquad \nonumber \\
\chi_3^{(1)}(\omega,T) & \approx & \frac{\epsilon_0 (\Delta \chi_1)^2 a^3}{k_BT} [N_{corr}(T)]_{av} {\cal K}\left(\omega \tau_{\alpha}\right).\qquad
\end{eqnarray}
Here $k_B$ is the Boltzman constant, and $\tau_{\alpha}(T)$ is the typical relaxation time at temperature $T$ corresponding to the 
relaxation frequency $f_{\alpha} = 1/(2 \pi \tau_{\alpha})$ where the imaginary part of the linear response is maximum. 
$\Delta \chi_1$ = $\chi_{1}(\omega = 0) - \chi_{1}(\omega \rightarrow \infty)$ is the part of the 
static linear susceptibility corresponding to the slow relaxation process of interest, $a^3$ is the volume occupied by one molecule, and 
${\cal H}$ and ${\cal K}$ are two complex scaling functions that approach zero for both small and large $\omega \tau_{\alpha}$. 
Note that the humped shapes of $|{\cal H}\left(\omega \tau_{\alpha}\right)|$ and $|{\cal K}\left(\omega \tau_{\alpha}\right)|$ are distinctive features of the glassy correlations.

BB's prediction relies on very general grounds, such as a generalised fluctuation dissipation relation, and was 
inspired by spin glass physics \cite{Lev88}, where a true second order phase transition happens at $T_c$, accompanied  by a critical divergence of
 $\chi_3$ (while the linear suceptibility $\chi_1$ does not diverge). A consequence of this generality is that the detailed expressions of the scaling
  functions $\cal H$ and $\cal K$ remain unkown. Here we present a phenomenological ``toy'' model where, for the first time, BB's predictions are 
   recovered with an explicit expression for the functions $\cal H$ and $\cal K$. By using plausible values of free parameters, the most
    salient experimental features of \cite{Cra10,Bru11} can be accounted for. Moreover we obtain new predictions on  higher order 
    nonlinear susceptibilities $\chi_{2k+1\ge 5}$. This could motivate new experiments deepening our understanding of the glass transition. 

\begin{figure}[t] 
\includegraphics*[width=8.0cm,height=6.0cm,angle=0]{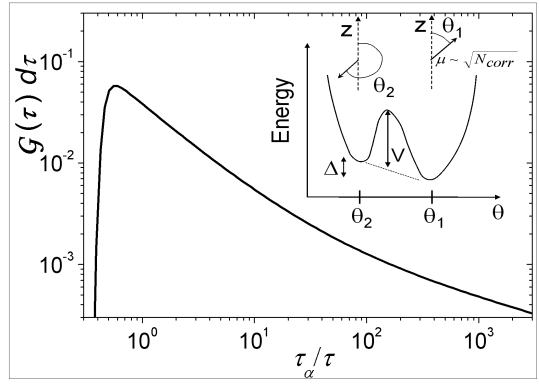}
\caption{\textit{Inset}: ADW model where each D.H. of $N_{corr}$ molecules evolves in an asymmetric double well making an angle $\theta_1$ with respect to the applied field $E$. \textit{Main graph}: distribution of relaxation times for Glycerol at $T=204.7$K, \cite{Blo03,Epa12}.} 
\label{fig1}
\end{figure}

\textit{Model}: We assume that all D.H.'s are independent from each other and that a given D.H. is a group of $N_{corr}$ molecules evolving in 
an Asymmetric Double Well potential (ADW), depicted in Fig. 
 \ref{fig1}. Each ADW is characterised by the height of its barrier $V$ and by an asymmetry energy $\Delta$. We neglect internal field effects. 
 On Fig. \ref{fig1}, $z$ represents 
 the axis of the external electric field $E(t)=E \cos(\omega t)$, and $\theta_1$ is the angle between the field and the well which has the 
  deepest energy at $E=0$. For simplicity we assume that $\theta_2 = \theta_1 + \pi$. With respect to earlier versions \cite{Wag99,Boh03}, 
  a key refinement is the assumption that the magnitude of the net dipolar moment $\mu$, in either of the two wells, is given by 
  $\mu = \mu_{molec}\sqrt{N_{corr}}$ where $\mu_{molec}$ is the molecular moment: This estimator of $\mu$ is assumed here because there 
  should not exist any geometrical ordering among the molecules contributing to a given D.H. \cite{Hur95}. 
  With $v_{DH} = N_{corr}a^3$ the volume of a D.H., the simplest approach, for $\theta_1=0$ and $\Delta=0$, 
  yields a static polarisation given by $(\mu/v_{DH})\tanh(\mu E/k_BT)$. Expanding in $E$ gives 
  $\Delta \chi_1 \propto \mu^2/v_{DH}$, which is independent of $N_{corr}$ since the $N_{corr}$ dependence of $\mu^2$ cancels that of $v_{DH}$. 
  For all higher orders such a cancellation does not happen, e.g. $\chi_3 \propto \mu^4/v_{DH} \propto N_{corr}$. This is the main reason why we 
  find below that $\chi_1$ is blind to $N_{corr}$ contrarily to all higher order susceptibilities.

Let us now consider a set of $N$ identical ADW's. With $\Pi_{j,k}$ the transition rate from the well $k$ to the well $j$, we obtain the number $n_1$ 
 -resp. $n_2$- of ADW's in state $1$ -resp. state $2$- by solving the two master 
 equations: $\partial n_1 / \partial t = -\Pi_{21}n_1 + \Pi_{12}n_2$ and 
$\partial n_2 / \partial t = -\Pi_{12}n_2 + \Pi_{21}n_1$. Assuming thermally activated barrier hoppings, one gets \cite{Wag99,Boh03}: $\Pi_{12} =  W \exp[(\Delta/2 + \mu E \cos\theta_1)/{k_B T}]$, 
 $\Pi_{21} = W \exp[ -(\Delta/2 + \mu E \cos\theta_1)/{k_B T}]$ 
where $W = \nu_0 \exp[-{V}/{k_B T}]$. Here $\nu_0=1/\tau_{m}$ where $\tau_{m}$ is the microscopic characteristic time of the thermal 
fluctuations within each well. The polarisation $P$ of the set of $N$ identical ADW's is given by $P = \mu \cos(\theta_1)(n_1-n_2)/(Nv_{DH})$. The two master equations yield the dynamical equation for $P$, which involves 
the relaxation time $\tau = 2W \cosh(\Delta/2k_BT)$ of the identical ADW's: 

\begin{eqnarray}
\tau \frac{dP}{dt} + P\left( \delta \sinh e+ \cosh e \right) &=&  {\cal M} \left( \delta \cosh e+ \sinh e \right) \nonumber \\
\hbox{where}\ e(t) \equiv F\cos(\omega t) \hbox{\ , }\ F&=& \frac{\mu_{molec} \sqrt{N_{corr}} \cos(\theta_1)}{k_B T} E \nonumber \\
\hbox{and}\ {\cal M}=\frac{\mu_{molec} \cos{\theta_1}}{\sqrt{N_{corr}} a^3} & \hbox{\ ,\ }& \delta = \tanh(\frac{\Delta}{2k_BT}). \ \ \ 
\label{eq3}
\end{eqnarray}

Setting $E=0$ in Eq. \ref{eq3} yields $P=P_0={\cal M}\delta$. As ${\cal M} \sim \cos \theta_1$, we obtain $<P_0>=0$ where the brackets denote 
the average over the isotropically distributed values of $\theta_1$. In the limit of small fields (i.e., $e \to 0$),  expanding Eq. \ref{eq3} to the 
first order in $e$ yields:

\begin{equation}
<P_1(t)> = \frac{<{\cal M}F> (1-\delta^2)}{\sqrt{1+(\omega \tau)^2}} \cos(\omega t -\arctan{\omega \tau}),
\label{eq4}
\end{equation}
i.e., a Debye response, as expected in any double well model \cite{Fro58}. As the linear dielectric spectra of supercooled liquids are asymmetric 
in frequency, we assume, as in other phenomenological models \cite{Ric06,Bru11b}, that the values of $\tau$   
are distributed according to ${\cal G}(\tau)$, \cite{Blo03}. ${\cal G}(\tau)$, given in \cite{Epa12} and in Fig. \ref{fig1}, is chosen to 
recover accurately the experimentally well known linear susceptibility $\chi_{1}(\omega, T)$ by weighting Eq. \ref{eq4} with ${\cal G}(\tau)$ 
and summing over all values of $\tau$. More precisely, ${\cal G}(\tau)$ determines the shape of $\chi_{1}(\omega, T)$,
 but not its overall magnitude $\Delta \chi_1$. We use 
the experimentally well known value of $\Delta \chi_1$ as an additional constraint in our model: from Eq. \ref{eq4},
 we obtain $\mu_{molec}^2 = 3k_BT\epsilon_0 a^3 \Delta \chi_1/(1-\delta^2)$; i.e., $\mu_{molec}$ is no longer a free parameter. 
 
 \begin{figure}[t]
\includegraphics*[width=9.0cm,height=6.0cm,angle=0]{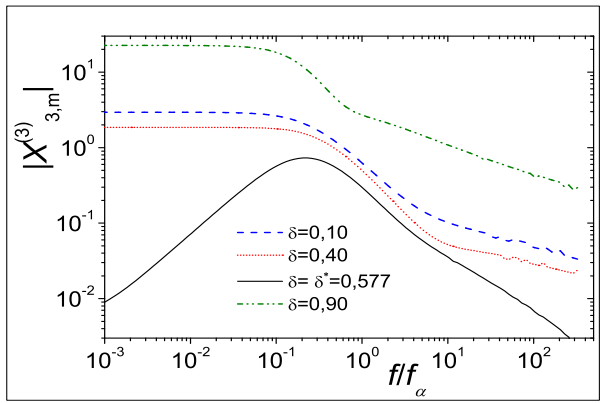}
\caption{(Color Online)  Effect of the dimensionless asymmetry energy $\delta$ on $\vert X_{3,m}^{(3)} \vert$: the spectra have a low pass character excepted close to $\delta^{\star} =1/\sqrt{3}$. Here $[N_{corr}]_{av}=5$.}
\label{fig2}
\end{figure}
 
 \textit{Computing} $\chi_3^{(3)}$ \textit{and} $\chi_3^{(1)}$: For a given value of $N_{corr}$ and of $\delta$, we consider all ADW's having 
 the same $\tau$ and $\theta_1$. By using Eq. \ref{eq3}, we compute the polarisation up to the third order in field. We first average 
 the result over $\theta_1$, then sum over $\tau$ with weight ${\cal G}(\tau)$, and finally average over the values of 
 $N_{corr}$ existing among various D.H.'s. The latter average is denoted by $[ \ ]_{av}$. This yields quantities -labelled below by an index ``$m$'' standing 
 for ``model''- which are comparable to experiments \cite{Cra10,Bru11}. Note that this method, by using the values of $a^3,\Delta \chi_1, {\cal G}(\tau)$ drawn from 
 standard experiments, eliminates $\theta_1, \nu_0, V$ and  $\mu_{molec}$. Thus, when comparing our model to the experimental values of 
 $\chi_3^{(3)}$ and $\chi_3^{(1)}$ at a given $T$, the two remaining free parameters are $[N_{corr}]_{av}$ and $\delta$. For simplicity we take a single value for $\delta$, and postpone the possible averaging over $\delta$ to Ref. \cite{Epa12}.
 
 In practice, we solve Eq. \ref{eq3} by assuming $e \ll 1$, and develop Eq. \ref{eq3}, as well as $P$, in series of $e$, up to the third order. 
  As the polarisation of a given set of ADW's sharing the 
 same $\theta_1,\tau,N_{corr}$ is not symmetric with respect to field reversal $E \to -E$, we  
 set $P(t)=\sum_{q=0}^{q=3} P_q(t)$ where $P_q\propto E^q$. Since $e \ll 1$, one has $\vert P_q \vert  \gg \vert P_{q'>q} \vert$; i.e., all $P_{q'>q}$ can
  be neglected when looking for $P_q$. Thus, $P_q$ is obtained by keeping only the terms $\propto e^q$ 
  in Eq. \ref{eq3}. This was illustrated above 
  to get first $P_0$ and then $P_1(t)$ -see Eq. \ref{eq4}-. Repeating the procedure to the order $e^2$ yields $P_2$, \cite{Epa12}. Finally going to the order $e^3$ gives:
 
\begin{equation}
\tau \frac{d(P_3)}{dt} + P_3 = \frac{{\cal M}(1- \delta^2)}{6} e^3  - \frac{1}{2} P_1 e^2- \delta P_2 e.
\label{eq5}
\end{equation}

\begin{figure}[t]
\includegraphics*[width=9.0cm,height=6.0cm,angle=0]{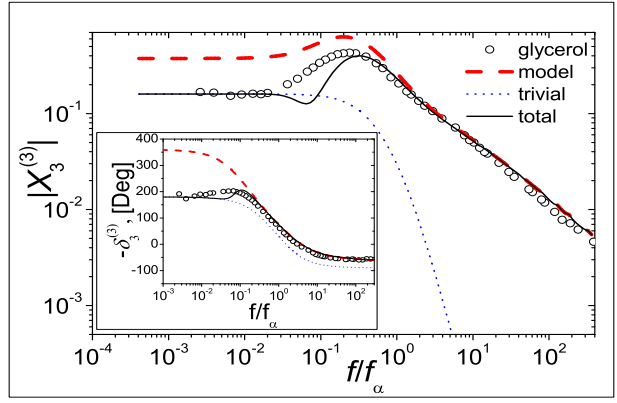}
\caption{(Color Online) For $[N_{corr}]_{av}= 5, \delta=0.60$ comparison of the ADW model with the experiments of  Ref. \cite{Bru11} at $T=204.7$K. 
$X_{3,tot}^{(3)}$ is the weighted sum (see text) of $X_{3,m}^{(3)}$ (see Eq. \ref{eq8}) and of $X_{3,trivial}^{(3)}$ corresponding to the 
cubic susceptibility of independant molecules undergoing rotational brownian motion \cite{Dej00,Bru11}. For $f/f_{\alpha} \ge 1$ one 
has $X_{3,m}^{(3)} \simeq X_{3,tot}^{(3)}$ and the experiments are very well accounted for by the model. For $f/f_{\alpha} \leq 1$, only the global 
trends of the data are restored by the model with $f_{ex}/f_{\alpha} = 0.14$. \textit{Inset:} Phases corresponding 
to the main graph, same symbols.}
\label{fig3}
\end{figure}

As $P_1$ and $P_2$ are known, the analytical expression of $P_3(t)$ is readily obtained from Eq. \ref{eq5}. 
After averaging over $\theta_1, \tau, N_{corr}$, one obtains $P_{3,m}$ that must be identified with
 the third order term 
$\mathscr P_3$ of the experimental polarisation. As $e^3 \propto E^3(3/4 \cos(\omega t) + 1/4 \cos(3\omega t))$, we recall that $\mathscr P_3$ 
naturally defines the first and third harmonics cubic susceptibilities (with phases $-\delta_3^{(k)}, k=1,3$) as \cite{Thi08}: 

\begin{equation}
\frac{\mathscr P_3 (t)}{\epsilon_0} = \frac{3E^3}{4} \vert \chi_{3}^{(1)}\vert \cos(\omega t - \delta_{3}^{(1)}) + \frac{E^3}{4} 
 \vert \chi_3^{(3)} \vert \cos(3\omega t - \delta_3^{(3)}).
 \label{eq6}
\end{equation}

\textit{Results of the ADW model:} A dimensional analysis shows that our model yields $P_{q,m} \propto [<{\cal M}F^q>]_{av}$, 
which has two important consequences.
 First it yields $P_{q,m} \propto <(\cos \theta_1)^{q+1}>$. This implies that the even terms $P_{2k,m} \equiv 0$, which ensures that the 
macroscopic polarisation reverses exactly upon the $E(t) \to -E(t)$ reversal, as required by macroscopic symmetry considerations. Second, 
 all odd terms $P_{2k+1,m}$ are non zero, yielding for the susceptibilities: $\chi_{2k+1,m} \propto \left[ N_{corr}^{k} \right]_{av}$. 
 This shows that the linear susceptibility $\chi_{1,m}$ is blind to the value of $N_{corr}$, contrary to
  higher order susceptibilities which are directly proportionnal to the $k^{th}$ moment
 of $N_{corr}$. This first important result is reminiscent of the spin-glass transition \cite{Lev88} which has inspired BB's prediction.

\begin{figure}[t]
\includegraphics*[width=9.0cm,height=6.0cm,angle=0]{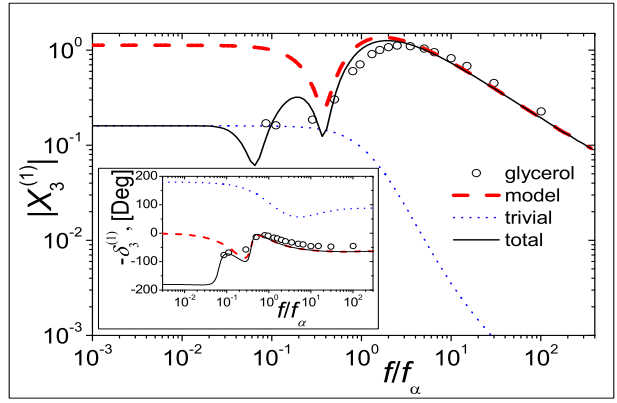}
\caption{(Color Online) Same symbols as in Fig. \ref{fig3}, excepted that $X_{3}^{(1)}$ is displayed here, and that $[N_{corr}]_{av}=15$. $\delta$ and $f_{ex}$ have the same values as in Fig. \ref{fig3}.}
\label{fig4}
\end{figure}

The above mentionned analysis yields $\chi_{3,m}^{(3)}$ that we convert into its dimensionless 
form $X_{3,m}^{(3)} = \chi_{3,m}^{(3)} k_BT/[\epsilon_0 a^3 (\Delta \chi_1)^2]$. Writing 
$X_{3,m}^{(3)} = \vert X_{3,m}^{(3)} \vert \exp[-i \delta_{3,m}^{(3)}]$ where $i^2=-1$, we get finally \cite{Epa12}, with  $x=\omega \tau$:

\begin{eqnarray}
X_{3,m}^{(3)} = \frac{9 [N_{corr}]_{av}}{5 (1-\delta^2)} \int\limits_{0}^{\infty}{{\cal G}(\tau)\frac{D_3^{(3)}(x)  
e^{i \left[\Psi_3^{(3)}(x)-\arctan(3x)\right]}}{\sqrt{1+(3x)^2}} d\tau } \nonumber \\
D_3^{(3)}(x) e^{i\Psi_{3}^{(3)}(x)} \equiv  \frac{1}{6} -\frac{e^{-i \arctan x}}{\sqrt{1+x^2}} \left[\frac{1}{2}-
\delta^2 \frac{e^{-i \arctan(2x)}}{\sqrt{1+4x^2}} \right]. 
\label{eq8}
\end{eqnarray}
 
 Note that $\cal G$ 
 nearly obeys Time-Temperature Superposition (TTS); i.e., it is nearly \cite{Bru11b} independent on $T$ when plotted as a function of 
 $\lambda = \tau/\tau_{\alpha}$. As $\omega \tau = \lambda \omega \tau_{\alpha}$, Eq. \ref{eq8} shows that $X_{3,m}^{(3)}$ 
 equals $[N_{corr}]_{av}$ times a function which does \textit{not} depend on $T$ -we take $\delta$ as a constant in $T$-, when plotted as a 
 function of $\omega \tau_{\alpha}$. Thus Eq. \ref{eq8} gives the 
 first phenomenological expression of the function ${\cal H}(\omega \tau_{\alpha})$ 
of Eq. \ref{eq1} -we recall that according to BB's prediction $X_{3}^{(3)}$ is $[N_{corr}]_{av}{\cal H}$-. Eq. \ref{eq8} thus shows 
explicitly that the $T$ dependence of $X_3^{(3)}$ is directly that of $[N_{corr}]_{av}$, up to small effects coming from small violations of 
TTS in ${\cal G}(\tau)$. This is the second important result of our model.

 Fig. \ref{fig2} shows the frequency behavior of $\vert X_{3,m}^{(3)} \vert$. For most values of $\delta$, 
 the spectrum has a low pass character. In the vicinity of $\delta^{\star}=1/\sqrt{3}$ the spectrum has a humped shape. To understand this, 
 let us note $P^{stat}$ the solution of Eq. \ref{eq3} at $\omega = 0$. One gets $P^{stat}= {\cal M} \tanh [e+\Delta/(2k_BT)]$. Expansion to 
 order $e^3$ yields $X_{3,m}^{stat}(\delta^{\star})=0$. Around $\delta^{\star}$, $X_{3,m}^{stat}$ moves from a 
 negative ``Ising-like'' value (low $\delta$'s), to a positive value for very asymmetric ADW's (high $\delta$'s). When $\omega \neq 0$, the
  effective relaxation time $\tau/(\cosh e+\delta \sinh e)$ comes into play, which contributes also to $X_{3,m}^{(3)}$. 
  This is why close to $\delta^{\star}$, $\vert X_{3,m}^{(3)} \vert$ has a humped shape in frequency. A deeper, i.e., much less model dependent, reason for this humped shape is given below. 

To compare our model to the nonlinear susceptibilities of glycerol reported in \cite{Cra10,Bru11}, we first focus on the case $f \ge f_{\alpha}$. 
Fig. \ref{fig3} shows that choosing $[N_{corr}]_{av}=5$ and $\delta = 0.60$ yields a very good agreement between our model and 
the values of $X_{3}^{(3)}(f\ge f_{\alpha})$ measured at $T=204.7$K $\simeq T_g+16$K. We emphasize that the agreement is good for both the 
modulus and the phase of $X_{3,m}^{(3)}$. Fig. \ref{fig4} shows the same kind of comparison for $X_3^{(1)}$, for which an expression similar to 
Eq. \ref{eq8} is given in \cite{Epa12}. On Fig. \ref{fig4} the best agreement between our model and the data reported in \cite{Bru11} 
is obtained with $[N_{corr}]_{av}=15$ and $\delta = 0.60$: With respect to the data, our model underestimates the phase by $\simeq 20^{\circ}$ 
and yields a maximum for the modulus at $f^{\star} \simeq 1.6 f_{\alpha}$ not far from the experimental value of $2.5f_{\alpha}$. The fact that 
the optimal $[N_{corr}]_{av}$ is not the same in Fig. \ref{fig3} and in Fig. \ref{fig4} may come from interferences between the nonlinear responses 
of the D.H.'s with different $\tau$, see Eq. \ref{eq8} and Ref. \cite{Epa12}. These interferences have different effects on $X_{3,m}^{(3)}$ 
and on $X_{3,m}^{(1)}$, see \cite{Epa12}, and this is not fully captured by our toy model, due to its simplicity. 
We emphasize, on the other hand, that $[N_{corr}]_{av}=5-15$ 
 is the right order of magnitude when comparing to the values given by $4$D-NMR experiments \cite{Tra98} or by Ref. \cite{Dal07}. 
 Moreover our model accounts for the fact that $\vert X_{3}^{(3)} \vert$ is peaked at a frequency ten times smaller than $\vert X_{3}^{(1)} \vert$
  in glycerol. Finally, $\delta=0.60$ amounts to $\Delta \simeq 1.4 k_BT \approx 1.4k_BT_g$, i.e. it does not introduce
  a new energy scale.

Now, let us move to the case $f<f_{\alpha}$. Here we must take into account the finite lifetime $\tau_{ex}$ of D.H.'s; i.e.,  
the fact that the liquid flows at large times \cite{Ric02}. The effective value of $[N_{corr}]_{av}$ decreases with frequency when 
$f \tau_{ex} \le 1$, since a given molecule is involved in various DH's at large times; i.e., it becomes independent of other molecules 
in the long run: as $X_{3,m}^{(k=1,3)} \propto [N_{corr}]_{av}$, \textit{this will give a humped shape to the nonlinear susceptibility even 
for the values of $\delta$ where $\vert X_{3,m}^{(k)} \vert$ has a low pass character}. 
To take this idea into account, we simply use the -well known- nonlinear response $X_{3,trivial}^{(k)}$ of 
independent molecules (see \cite{Dej00,Cra10,Bru11}) and assume that 
it dominates the measured $X_{3}^{(k)}$ when $f \tau_{ex} \ll 1$. In practice, we write heuristically the total cubic 
susceptibility $X_{3,tot}^{(k)}$ (with once again $k=1,3$) as: $X_{3,tot}^{(k)} = pX_{3,m}^{(k)}+ (1-p)X_{3,trivial}^{(k)}$ with 
$p=\exp{(-f_{ex}/f)}$, see \cite{Epa12}. For $f>f_{\alpha}$, $X_{3,tot}^{(k)}$ is 
of course very close to $X_{3,m}^{(k)}$, since $p \simeq 1$. For $f \le f_{\alpha}$, Figs. \ref{fig3}-\ref{fig4} show that, 
with $f_{ex} = 0.14f_{\alpha}$, $X_{3,tot}^{(k)}$ has the same global qualitative trends as the measured 
$X_{3}^{(k)}$ in glycerol. We note that $f_{ex}/f_{\alpha} = 0.14 $ amounts to $Q=\tau_{ex} / \tau_{\alpha} \simeq 7$, 
which is compatible with the values $Q \simeq 3-10$ reported before \cite{Ric02} albeit still debated \cite{Epa12}. We think that the oscillation of 
$\vert X_{3,tot}^{(k)}\vert$ around $0.1f_{\alpha}$ is unphysical and comes from the very naive 
way of including $\tau_{ex}$ in our analysis. 

To conclude, we have developped a very simple toy model for the nonlinear susceptibilities in supercooled liquids.
 We find that $\chi_{2k+1,m} \propto [N_{corr}^{k}]_{av}$; i.e., that 
 $\chi_{1,m}$ is blind to the value of $N_{corr}$ contrary to all higher order susceptibilities. This 
 yields the first phenomenological expression of the scaling functions involved in BB's predictions. With  
  reasonnable values of parameters, the main trends of nonlinear experimental data are recovered. Our model explains very simply 
  why the nonlinear responses yield brand new information on the glassy dynamics. 
  This simplicity may trigger more experiments deepening our understanding of the glass transition.

We thank R. Tourbot for his outstanding technical help, S. Nakamae for carefully reading the paper.  
We thank G. Diezemann for encouraging discussions in dec. 2011, and C.Alba-Simionesco, G. Biroli, J.-P. Bouchaud, J.-P. Carton, 
P.M. D\'ejardin for long lasting help.

{\bf Note added: See also on the same subject the paper of Gregor Diezemann to appear on Condmatt TODAY plus or minus a few days !...}

\renewcommand{\figurename}{Fig. S}
\par
\begin{center}
\vskip 10mm

{\bf Electronic Physics Auxiliary Publication Service: Supplementary Information for:}

\par
\vskip 10mm

	\textbf{\large{Nonlinear dielectric susceptibilities in supercooled liquids: a toy model.}}
	\par
	\null
	\par
	 F. Ladieu, C. Brun, D. L'H\^ote

\end{center}
\vskip 10mm

We detail hereafter the calculations for $X_{3,m}^{(3)}$ and $X_{3,m}^{(1)}$ summarized in our main article. We then give  a short justification of our assumption $p=\exp[-f_{ex}/f]$ made in the end of the main article. Finally, we give more informations about what happens when averaging over the dimensionless asymmetry parameter $\delta$.

\section{The nonlinear susceptibilities in the Asymmetric Double Well Potential model.}
\subsection{Calculations for one set of identical Asymmetric Double Wells.}
 
 In this section we consider a set of $N$ identical Asymmetric Double Wells (ADW); i.e., a set of ADW's sharing the same values for  all microscopic parameters of the model. 
 Denoting $n_{1}$ (respectively $n_2$) the number of ADW's in state $\theta_1$ (respectively $\theta_2=\theta_1+\pi$), the polarisation $P$ of 
 the considered set of ADW's is given by: 

\begin{equation}
P = \frac{(n_1-n_2) \mu_{molec} \sqrt{N_{corr}} \cos{\theta_1}}{N N_{corr}a^3} = {\cal M}\frac{n}{N} \ \hbox{\ where\ }\ {\cal M}=\frac{\mu_{molec} \cos{\theta_1}}{\sqrt{N_{corr}} a^3},
\label{eqS1}
\end{equation}
where  it was assumed that the net dipolar moment in either of the two states of a given ADW is given by 
$\mu =\mu_{molec} \sqrt{N_{corr}}$, with $\mu_{molec}$ the molecular dipole moment (see the main article). Combining the two master equations for $n_{1}(t)$ and $n_{2}(t)$, with $n_{2}(t) = N-n_{1}(t)$, one gets 
the equation for the dynamics of $P$: 

\begin{eqnarray}
\tau \frac{dP}{dt} + P\left( \delta \sinh e+ \cosh e \right) &=&  {\cal M} \left( \delta \cosh e+ \sinh e \right) \nonumber \\
\hbox{where}\ e(t) \equiv F\cos(\omega t) \hbox{\ , }\ F&=& \frac{\mu_{molec} \sqrt{N_{corr}} \cos(\theta_1)}{k_B T} E \nonumber \\
\hbox{and}\ {\cal M}=\frac{\mu_{molec} \cos{\theta_1}}{\sqrt{N_{corr}} a^3} & \hbox{\ ,\ }& \delta = \tanh(\frac{\Delta}{2k_BT}). \ \ \ 
\label{eqS2}
\end{eqnarray}

As explained in the article, the two sources of nonlinearity in Eq. \ref{eqS2} are: \textit{(i)} the nonlinear character of 
the equilibrium value $P^{stat}={\cal M}( \delta \cosh e+ \sinh e )/(\delta \sinh e+ \cosh e )  = {\cal M} \tanh [e+\Delta/(2k_BT)]$; and 
\textit{(ii)} the nonlinear character of the instantaneous relaxation time $\tau_{eff} = \tau/(\delta \sinh e+ \cosh e)$. 

We expand $P(t)$ in series of powers of the field $E$ up to third order $P(t)=P_0 + P_1(t)+P_2(t)+P_3(t)$ where $P_{q} \propto E^q$. As 
$E(t)^2 = E^2(1+\cos(\omega t))/2$ and $E(t)^3 = E^3(3\cos(\omega t)+\cos(3\omega t))/4$, $P_2(t)$ and $P_3$ are the sum of two 
terms: 

\begin{eqnarray} 
P_2(t)=P_2^{(0)} + P_2^{(2)}(t) \nonumber \\
P_3(t)=P_3^{(1)}(t) + P_3^{(3)}(t) , 
\label{eqS3}
\end{eqnarray}
where the superscript in parentheses indicates the index of the relevant harmonics. For example, $P_3(t)$ is given by a term oscillating at the 
fundamental frequency, and by a term oscillating at three times the fundamental frequency. 

As the condition $e\ll 1$ is well obeyed experimentally, one can neglect all  $P_{q'>q}$ terms when computing $P_q$.
 Therefore $P_q$ is obtained by keeping only the terms $\propto e^q$ in Eq. \ref{eqS2} above.

To the order $e^0$ it is found that:

\begin{equation}
P_0 = {\cal M} \delta .
\label{eqS4}
\end{equation}

Now, going to the order $e^1$, one has (by using the result for $P_0$ in Eq. \ref{eqS4}): 

\begin{equation}
\tau \frac{dP_1}{dt} + P_1 =  {\cal M} \big[1- \delta^2 \big] \times F\cos(\omega t) , 
\label{eqS5}
\end{equation}
which yields: 

\begin{equation}
P_1(t) = \frac{{\cal M}(1-\delta^2)}{\sqrt{1+(\omega \tau)^2}} F \cos\left(\omega t - \arctan(\omega \tau)\right) .
\label{eqS6}
\end{equation}

We now go to the order $e^2$ and get:

\begin{equation}
\tau \frac{d(P_2)}{dt} + P_2 = - \delta F P_1(t) \cos(\omega t) .
\label{eqS7}
\end{equation}

As $P_1(t)$ oscillates at frequency $\omega$, the right hand side of Eq. \ref{eqS7} contains one constant term and another term oscillating at 
$2\omega$. Therefore, one finds:  

\begin{eqnarray}
P_{2}^{(0)} & = & \frac{{\cal M}( \delta - \delta^3)F^2}{2 \sqrt{1+(\omega \tau)^2}} \cos\left[\pi+\arctan(\omega \tau)\right] \nonumber \\
P_{2}^{(2)}(t) & = & \frac{{\cal M}( \delta - \delta^3)F^2}{2 \sqrt{1+(\omega \tau)^2}\sqrt{1+(2\omega \tau)^2}} \cos\left[2 \omega t +\pi 
- \arctan(\omega \tau)- \arctan(2 \omega \tau)\right] .
\label{eqS8}
\end{eqnarray}

Finally, we reach the order $e^3$ and get: 

\begin{equation}
\tau \frac{d(P_3)}{dt} + P_3 = (1/6) {\cal M}(1- \delta^2)[e(t)]^3 -P_2(t) \delta e(t) - P_1(t) [e(t)]^2/2 .
\label{eqS9}
\end{equation}

We separate the terms oscillating at $\omega$ from  those oscillating at $3\omega$. Denoting by $\vert P_1 \vert$ -respectively $\vert P_2^{(2)} \vert$- the amplitude of $P_1(t)$ -respectively $P_2^{(2)}(t)$-, one obtains:

\begin{eqnarray}
\tau \frac{d(P_3^{(1)})}{dt} + P_3^{(1)} & = & (1/8){\cal M}F^3(1-\delta^2) \cos(\omega t) - (1/4)F^2 \vert P_1\vert \Big[ \cos[\omega t - \arctan(\omega \tau)] \nonumber \\
\ &\ & +(1/2) \cos[\omega t + \arctan(\omega \tau)] \Big] -P_2^{(0)}\delta F \cos(\omega t)- \nonumber \\
\ &\ & - (1/2)\vert P_2^{(2)} \vert \delta F \cos[\omega t+ \pi -\arctan(\omega \tau)-\arctan(2\omega \tau)] .
\label{eqS10}
\end{eqnarray}
as well as

\begin{eqnarray}
\tau \frac{d(P_3^{(3)})}{dt} + P_3^{(3)} & = & (1/24){\cal M}F^3(1-\delta^2) \cos(3\omega t) - (1/8)F^2 \vert P_1\vert  \cos[3\omega t - \arctan(\omega \tau)] \nonumber \\
\ &\ & - (1/2)\vert P_2^{(2)} \vert \delta F \cos[3\omega t+ \pi -\arctan(\omega \tau)-\arctan(2\omega \tau)] .
\label{eqS11}
\end{eqnarray}

By using Eqs. \ref{eqS6}-\ref{eqS8} and the two previous equations, one finds: 

\begin{eqnarray}
\tau \frac{d(P_3^{(1)})}{dt} + P_3^{(1)} & = & 
(1/4){\cal M}F^3(1-\delta^2) D_3^{(1)}(\omega \tau)\cos\left[\omega t + \Psi_3^{(1)}(\omega \tau)\right] \nonumber \\
\ {\hbox{where}}\ D_3^{(1)}(\omega \tau)\cos(\omega t + \Psi_3^{(1)}(\omega \tau)) & \equiv & \left(\frac{1}{2}+
 \frac{2 \delta^2}{1+(\omega \tau)^2} \right)\cos(\omega t)- \frac{\cos\left[\omega t - \arctan(\omega \tau)\right]}
 {\sqrt{1+(\omega \tau)^2}} \nonumber \\
\ &\ & - \frac{\cos\left[\omega t + \arctan(\omega \tau)\right]}{2\sqrt{1+(\omega \tau)^2}} \nonumber \\
\ &\ & - \frac{\delta^2 \cos\left[\omega t +\pi - \arctan(\omega \tau)- \arctan(2\omega \tau)\right]}{\sqrt{1+(\omega \tau)^2}
\sqrt{1+(2\omega \tau)^2}} ,
\label{eqS12}
\end{eqnarray}
as well as

\begin{eqnarray}
\tau \frac{d(P_3^{(3)})}{dt} + P_3^{(3)} & = & (1/4){\cal M}F^3(1-\delta^2) D_3^{(3)}(\omega \tau)\cos\left[3\omega t + \Psi_3^{(3)}(\omega \tau)\right] \nonumber \\
\ {\hbox{with}}\ D_3^{(3)}(\omega \tau)\cos(3\omega t + \Psi_3^{(3)}(\omega \tau)) & \equiv & (1/6)\cos(3\omega t)- \frac{\cos\left[3 \omega t - \arctan(\omega \tau)\right]}{2 \sqrt{1+(\omega \tau)^2}} \nonumber \\
\ &\ & - \frac{\delta^2 \cos\left[3 \omega t +\pi - \arctan(\omega \tau)- \arctan(2\omega \tau)\right]}{\sqrt{1+(\omega \tau)^2}\sqrt{1+(2\omega \tau)^2}} .
\label{eqS13}
\end{eqnarray}
Note that the above definitions of $D_{3}^{(3)}$ and $\Psi_3^{(3)}$ are consistent with those in the main article. 

The solution of Eq. \ref{eqS12} is given by

\begin{equation}
P_3^{(1)}(t) = \frac{{\cal M}(1-\delta^2)F^3}{4 \sqrt{1+(\omega \tau)^2}}D_3^{(1)}(\omega \tau)\cos\left[\omega t + \Psi_3^{(1)}(\omega \tau)-\arctan(\omega \tau)\right] .
\label{eqS14}
\end{equation}

The solution of Eq. \ref{eqS13} is given by 
\begin{equation}
P_3^{(3)}(t) = \frac{{\cal M}(1-\delta^2)F^3}{4 \sqrt{1+(3\omega \tau)^2}}D_3^{(3)}(\omega \tau)\cos\left[3\omega t + \Psi_3^{(3)}(\omega \tau)-\arctan(3\omega \tau)\right] .
\label{eqS15}
\end{equation}

\subsection{Averaging over $\theta_1$,$\tau$ and $N_{corr}$.}

As explained in the main article, three kinds of averages must be done in our ADWP model: 

\textit{(i)} First, we have to average over the angle $\theta_1$, the values of which are assumed to be isotropically distributed. Denoting
 this average by $<\ >$, one finds $<(cos \theta_1)^{2k+1}> =0$ 
 and $<(cos \theta_1)^{2k}> =1/(2k+1)$, for any integer $k$. As we have found above that 
 $P_q \propto {\cal M}F^q$, one obtains $<P_q> \propto <{\cal M}F^q> \propto <(cos \theta_1)^{q+1}>$. Therefore, all the even integer harmonics 
 vanish, contrarily to all odd harmonics which are found to be 
 
 \begin{equation}
 <P_{2k+1}> \propto N_{corr}^k .
 \label{eqS16}
 \end{equation}

\textit{(ii)} Second we have to average over various relaxation times $\tau$, with weight ${\cal G}(\tau) d\tau$. The distribution function $\cal G$ is chosen so as to recover accurately the experimental linear response, $\chi_1$. Therefore 
 $\cal G$ must simultaneously solve the two following equations for the real part, $\chi_1'$, and the imaginary part,  $\chi_1''$: 
 
\begin{eqnarray}
 \frac{\chi_{1}'(\omega)-\chi_{1}'(\infty)}{\Delta \chi_{1}} & = & \int\limits_{-\infty}^{\infty}{{\cal G}(\ln \tau)\times \frac{1}{1+(\omega \tau)^2} \times d\ln \tau} \nonumber \\
 \frac{\chi_{1}''(\omega)}{\Delta \chi_{1}} & = & \int_{- \infty}^{\infty} {{\cal G}(\ln \tau)\times \frac{\omega \tau}{1+(\omega \tau)^2} \times d\ln \tau,}
\label{eqS17}
\end{eqnarray}
where we have used the fact that ${\cal G}(\tau) d\tau = {\cal G}(\ln \tau) d\ln \tau$. In practice one uses \cite{Blo03,Bru11b}:

\begin{eqnarray}
 {\cal G}(\ln(\tau)) & = & N_{GGE} e^{-(\frac{\beta}{\alpha})\left(\frac{\tau}{\tau_0} \right)^\alpha}   \left(\frac{\tau}{\tau_0}\right)^{\beta}\left[1+\left(\frac{\sigma \tau}{\tau_0} \right)^{\gamma - \beta} \right] \nonumber \\
 {\hbox {\ with\ }} N_{GGE} & = & \frac{\alpha \left(\frac{\beta}{\alpha} \right)^{\frac{\beta}{\alpha}}}{\Gamma\left(\frac{\beta}{\alpha} \right)+\sigma^{\gamma-\beta} \left(\frac{\alpha}{\beta} \right)^{(\gamma-\beta)/{\alpha}} \Gamma\left(\frac{\gamma}{\alpha} \right)}.
\label{eqS18}
\end{eqnarray}

Here $\Gamma (x)$ is the Euler gamma function and $\alpha, \beta, \sigma, \gamma, \tau_{0}$ are $T$ dependent parameters. For glycerol, a good set of parameters is given by:

\begin{eqnarray}
\alpha & = & 10 \nonumber \\
\beta & = & -5.5996 \times 10^{-1} + 4.0900 \times 10^{-3} T + 1.50795 \times 10^{-5} T^2 \nonumber \\
\sigma & = & 1.57 \times 10^{-1}\exp\left[\frac{407.525}{T - 141}\right] \nonumber \\
\gamma & = & -7.826920 + 1.015 \times 10^{-1} T - 4.32345 \times 10^{-4} \times T^2 + 
  6.34415 \times 10^{-7} \times T^3 \nonumber \\
\tau_0 & = & 1.1511 \times 10^{-15} \times \exp\left[\frac{19.08905 \times 127.38588}{T - 127.38588}\right],
\label{eqS19}
\end{eqnarray} 
with $T$ expressed in Kelvins.

Note that $\tau_0$ is nearly proportionnal to the typical relaxation time $\tau_{\alpha}$ defined by $\tau_{\alpha} = 1/(2\pi f_{\alpha})$ 
where $f_{\alpha}$ is the frequency of the peak of $\chi_1''$. Additionally one finds from Eq. \ref{eqS17}:

\begin{equation}
\Delta \chi_1 = \frac{(1-\delta^2)(\mu_{molec})^2 <[\cos(\theta_1)]^2>}{k_BT \epsilon_0 a^3} , 
\label{eqS20}
\end{equation}
and with $<(\cos \theta_1)^2> = 1/3$, we obtain 

\begin{equation}
(\mu_{molec})^2 = \frac{3k_BT \epsilon_0 a^3\Delta \chi_1}{(1-\delta^2) } .
\label{eqS21}
\end{equation}

At this point the two free parameters of our ADWP model are $N_{corr}$ and $\delta$. In particular Eq. \ref{eqS21} sets the value of $\mu_{molec}$, since $\Delta \chi_1$ and $a^3$ are experimentally well known. 

\textit{(ii)} The third and the last average to be taken is over the values of $N_{corr}$. Indeed the proportionality expressed by Eq. \ref{eqS16} 
remains true when averaging over the $\tau$'s. Therefore, it is very easy in our model to take into account the fact that there exists a distribution of the values of $N_{corr}$ among various dynamical heterogeneities of a real supercooled liquid. As all above equations 
have been derived for given  free parameters $N_{corr}$ and $\delta$, we superpose the ensemble of models with the same $\delta$ but with 
different values of $N_{corr}$. Denoting the average over $N_{corr}$ by $[\ ]_{av}$, we obtain from Eq. \ref{eqS16}: 

 \begin{equation}
 [<P_{2k+1}>]_{av} \propto \chi_{2k+1,m} \propto [N_{corr}^k]_{av} ,
 \label{eqS22}
 \end{equation}
where $\chi_{2k+1,m}$ is the macroscopic nonlinear susceptibility of the order $2k+1$, and where the index $m$ stands for ``model'', so as to avoid 
any confusion between the nonlinear susceptibilities produced by the model and those corresponding to what is experimentally measured 
(denoted $\chi_{2k+1}$). Note that $\chi_{2k+1,m}$ represents generically the set of components of the 
macroscopic polarisation which is proportionnal to $E^{2k+1}$ and oscillates at one of the odd harmonics between $1\omega$ and $(2k+1)\omega$. 
For example, $\chi_{3,m}$ corresponds to two terms: one is proportionnal to $\chi_{3,m}^{(1)}$ -note the presence of exponent $(1)$- and oscillates at $1\omega$, and the other one is 
proportionnal to $\chi_{3,m}^{(3)}$ and oscillates at $3\omega$.

\subsection{Explicit expressions for the cubic susceptibilities $\chi_{3}^{(3)},\chi_{3}^{(1)}$.}

The macroscopic polarisation $\mathscr P$ is given by \cite{Thi08}: 

\begin{equation}
\frac {\mathscr P(t)}{\epsilon_0} =  
\int_{-\infty}^{\infty}{\chi_1(t-t')E(t')dt'} + \iiint_{-\infty}^{\infty} {\chi_3(t-t'_1,t-t'_2,t-t'_3) E(t'_1)E(t'_2)E(t'_3)dt'_1dt'_2dt'_3} + ..., 
\label{eqS22b}
\end{equation}
where the function $\chi_1(t)$ corresponds to the experimental macroscopic linear response 
while $\chi_3(t_1,t_2,t_3)$ is the experimental macroscopic nonlinear response.

It is shown in ref. \cite{Thi08}, that  for a field $E(t)= E \cos(\omega t)$ one gets: 

\begin{equation}               
\frac{\mathscr P(t)}{\epsilon_0} = E\left|\chi_1\right| \cos(\omega t - \delta_1) +3/4 E^3\left|\chi_{3}^{(1)}\right| \cos(\omega t - \delta_{3}^{(1)}) + 1/4E^3\left|\chi_3^{(3)}\right| \cos(3\omega t - \delta_3^{(3)})+...  
\label{eqS22c} 
\end{equation}

We now must identify the result of our model with  above relations giving the experimental macroscopic polarisation. 
We start from Eqs. \ref{eqS14}-\ref{eqS15} and average over $\theta_1$ which yields: 

\begin{equation}
<{\cal M}F^3> = \frac{N_{corr} (\mu_{molec})^4 \left< (\cos \theta_1)^4 \right>}{a^3 (k_B T)^3}E^3 = \frac{9N_{corr}}{5(1-\delta^2)^2} \frac{\epsilon_0^2 a^3 (\Delta \chi_1)^2}{k_BT}E^3.
\label{eqS23}
\end{equation}
We then average over the $\tau$'s, as in Eq. \ref{eqS17}, and then over $N_{corr}$. With Eqs. \ref{eqS15}-\ref{eqS22c}, we obtain:
 
\begin{eqnarray}
\int\limits_{-\infty}^{\infty}{ \frac{\left[ <{\cal M}F^3> \right]_{av}(1-\delta^2)}{4 \sqrt{1+(3\omega \tau)^2}}D_3^{(3)}(\omega \tau)\cos\left[3\omega t + \Psi_3^{(3)}(\omega \tau)-\arctan(3\omega \tau)\right] {\cal G}(\ln \tau) d\ln \tau} && \nonumber \\
\equiv  (1/4)\epsilon_0 E^3\left|\chi_3^{(3)}\right| \cos(3\omega t - \delta_3^{(3)}) && \nonumber \\
= (1/4)E^3 \frac{\epsilon_0^2 (\Delta \chi_1)^2 a^3}{k_B T} [N_{corr}]_{av}\vert {\cal H}(\omega \tau_{\alpha})\vert \cos(3\omega t + \arg({\cal H}(\omega \tau_{\alpha}))), && 
\label{eqS24}
\end{eqnarray}
where the last equality was obtained by replacing $\chi_3^{(3)}$ by Bouchaud-Biroli's prediction $\chi_3^{(3)} \approx \frac{\epsilon_0 (\Delta \chi_1)^2 a^3}{k_BT} [N_{corr}]_{av} \, {\cal H}\left(\omega \tau_{\alpha}\right)$, see the main article. Combining Eqs. \ref{eqS23}-\ref{eqS24}, one obtains: 

\begin{eqnarray}
&[N_{corr}]_{av}&\vert {\cal H}(\omega \tau_{\alpha})\vert \cos(3\omega t + \arg({\cal H}(\omega \tau_{\alpha}))) = \nonumber \\ 
&& \frac{9 [N_{corr}]_{av}}{5(1-\delta^2)} \int\limits_{-\infty}^{\infty}{{\cal G}(\ln \tau)   \frac{D_3^{(3)}(\omega \tau)\cos\left[3\omega t + \Psi_3^{(3)}(\omega \tau)-\arctan(3\omega \tau)\right]}{\sqrt{1+(3\omega \tau)^2}} d\ln \tau}.
\label{eqS25}
\end{eqnarray}
As in the main article one defines the dimensionless nonlinear suscceptibility as 
$X_{3,m}^{(3)} = \chi_{3,m}^{(3)} k_BT/[\epsilon_0 a^3 (\Delta \chi_1)^2]$. Writing 
$X_{3,m}^{(3)} = \vert X_{3,m}^{(3)} \vert \exp[-i \delta_{3,m}^{(3)}]$ one obtains: 

\begin{eqnarray}
\vert X_{3,m}^{(3)}\vert & = & \frac{9 [N_{corr}]_{av}}{5 (1-\delta^2)} \sqrt{\left({\cal S}^{(3)}_{COS}\right)^2+\left({\cal S}^{(3)}_{SIN}\right)^2} \nonumber \\
\ {\hbox{\ and \ }}\ -\delta_{3,m}^{(3)} & \equiv & {\hbox{\ phase \ of \ }}X_{3,m}^{(3)}  =  \arctan \left( \frac{{\cal S}^{(3)}_{SIN}}{{\cal S}^{(3)}_{COS}}\right) \nonumber \\
\ {\hbox{\ with \ }}\ {\cal S}^{(3)}_{COS} & = & \int\limits_{-\infty}^{\infty}{{\cal G}(\ln \tau) \frac{D_3^{(3)}(\omega \tau)\cos\left[\Psi_3^{(3)}(\omega \tau)-\arctan(3\omega \tau)\right]}{\sqrt{1+(3\omega \tau)^2}} d\ln \tau} \nonumber \\
\ {\hbox{\ and \ }}\ {\cal S}^{(3)}_{SIN}  & = & \int\limits_{-\infty}^{\infty}{{\cal G}(\ln \tau) \frac{D_3^{(3)}(\omega \tau)\sin\left[\Psi_3^{(3)}(\omega \tau)-\arctan(3\omega \tau)\right]}{\sqrt{1+(3\omega \tau)^2}} d\ln \tau}.
\label{eqS26}
\end{eqnarray}

A similar calculation for $\chi_{3,m}^{(1)}$ yields: 

\begin{eqnarray}
\vert X_{3,m}^{(1)}\vert & = & \frac{3 [N_{corr}]_{av}}{5 (1-\delta^2)} \sqrt{\left({\cal S}^{(1)}_{COS}\right)^2+\left({\cal S}^{(1)}_{SIN}\right)^2} \nonumber \\
\ {\hbox{\ and \ }}\ -\delta_{3,m}^{(1)} & \equiv & {\hbox{\ phase \ of \ }}X_{3,m}^{(1)}  =  \arctan \left( \frac{{\cal S}^{(1)}_{SIN}}{{\cal S}^{(1)}_{COS}}\right) \nonumber \\
\ {\hbox{\ with \ }}\ {\cal S}^{(1)}_{COS} & = & \int\limits_{-\infty}^{\infty}{{\cal G}(\ln \tau) \frac{D_3^{(1)}(\omega \tau)\cos\left[\Psi_3^{(1)}(\omega \tau)-\arctan(\omega \tau)\right]}{\sqrt{1+(\omega \tau)^2}} d\ln \tau} \nonumber \\
\ {\hbox{\ and \ }}\ {\cal S}^{(1)}_{SIN}  & = & \int\limits_{-\infty}^{\infty}{{\cal G}(\ln \tau) \frac{D_3^{(1)}(\omega \tau)\sin\left[\Psi_3^{(1)}(\omega \tau)-\arctan(\omega \tau)\right]}{\sqrt{1+(\omega \tau)^2}} d\ln \tau}.
\label{eqS27}
\end{eqnarray}
Note that in the first equality of Eq. \ref{eqS27} there is a factor $3$ instead of  $9$ found in Eq. \ref{eqS26}. This comes from  Eq. \ref{eqS22c} where there is a factor $3/4$ for the cubic term oscillating at $\omega$ while it is only $1/4$ for the cubic term oscillating at $3\omega$.

\section{More on the frequency dependence of the weight $p=\exp[-f_{ex}/f]$.}

All above calculations have been made as if the lifetime $\tau_{ex}$ of the considered Asymmetric Double Wells is infinite. 
As a supercooled liquid is ergodic above $T_g$, the heterogeneity of the dynamics implies that $\tau_{ex}$ must be finite. This comes from the fact that a region of 
space relaxing faster than the average must become a region relaxing slower than the average, to restore ergodicity. We shall assume, for 
simplicity, that any ADW is reshuffled with the same characteristic time $\tau_{ex}$, whatever the value of $\tau$ it had just before. 

After reshuffling, the glassy correlations are different from those established before. Thus, if one performs \textit{an average over time    
longer  than} $\tau_{ex}$, a given molecule is no longer correlated to any other molecule. This is why, one expects any molecule to become \textit{effectively} independent of all other molecules in the limit of large 
times $t \gg \tau_{ex}$. Therefore one expects, at large times, the measured nonlinear dimensionless susceptibilities $X_{3}^{(k)}$ to be 
dominated by the corresponding susceptibilities $X_{3,trivial}^{(k)}$ of independent molecules undergoing Brownian rotational motion. 
Note that $X_{3,trivial}^{(k)}$ has been fully calculated in Ref. \cite{Dej00}.

Very few things are quantitatively established concerning the reshuffling phenomenon; even the value of $Q=\tau_{ex}/\tau_{\alpha}$ remains a subject 
of discussions \cite{SHDbook08}. Therefore a detailed description of its impact on nonlinear susceptibilities is not available at present. This 
is why, we heuristically add the nonlinear susceptiblities $X_{3,m}^{(k)}$ given by our ADWP model (multiplied by the weight $p$) 
to $X_{3,trivial}^{(k)}$ (multiplied by the complementary weight $(1-p)$). As the limit of large times $t \gg \tau_{ex}$ corresponds to  
low frequencies $f \tau_{ex} \le 1$, we physically expect that $p$ vanishes in this limit. The simplest way to express this idea quantitatively is to state that  the weight $(1-p)$ of the trivial response is given by the probability of a reshuffling event happening during one $E$ oscillation period of $2 \pi /\omega$. It is reasonnable to assume that the probability of the reshuffling events are given by a 
Poissonian distribution $(1/\tau_{ex})\exp{(-t/\tau_{ex})}$, and therefore:

\begin{equation}
1-p = \int_{0}^{2\pi/\omega}{\exp{(-t/\tau_{ex})} \frac{dt}{\tau_{ex}}}  \ \hbox{\ which yields\ }\ p=\exp{(-f_{ex}/f)} \hbox{\ where \ } f_{ex} = 1/\tau_{ex} .
\label{eqS28}
\end{equation}

This is the weighting function that has been used in Figs. 3-4 of the main article. Of course it plays a role only for the range 
$f\le f_{\alpha}$ as one has $p(f \ge f_{\alpha}) \simeq 1$ since $f_{ex} /f_{\alpha} \ll 1$. 

\section{Averaging over $\delta$.}

For simplicity we have presented in the main article the results of our ADWP model obtained for a single value of the dimensionless asymmetry 
$\delta$. One can generalise the results by averaging  over $\delta$, at the cost of additional parameters. To investigate 
this question, we computed the values of $X_{3,m}^{(k)}$ for $100$ values of $\delta$ linearly distributed in the $[0; 0.99]$ interval. We 
then averaged the complex values of $X_{3,m}^{(k)}$ by a weight $w(\delta)$. For simplicity we have used either a flat distribution which 
is non zero only between $\delta_{min}\ge 0$ and $\delta_{max} < 1$; or a ``gaussian'' distribution where 
$w(\delta) = C\times \exp{\left[-(\delta - \delta_1)^2/(2\times (\delta_{2})^{2}) \right]}$. Here $C$ is the proper normalisation constant 
taking into account that $\delta$ is defined only on the $[0;1]$ interval. Note that $\delta_1$ is close to, but not exactly equal to, 
the average of $\delta$; and similarly $\delta_2$ is not exactly its standard deviation due to the fact that $\delta$ is restricted to the 
$[0;1]$ interval.

Two interesting features are worth noting in this averaging procedure over $\delta$:

$\bullet$ First, the values of $X_{3,m}^{(k)}$ plotted in Figs. 3-4 of the main article can be recovered with distributed values of $\delta$. 
For example, Fig. S \ref{Epaps-Fig1} below shows the values of $\delta_1$ and of $\delta_2$ that have to be chosen to recover the values of $X_{3,m}^{(k)}$ plotted in Figs. 3-4, by using a 
gaussian distribution. One sees in Fig. S \ref{Epaps-Fig1}, that the value $\delta=0.60$, chosen in 
the main article to fit the experiments without averaging over $\delta$, corresponds to the limiting case of a gaussian distribution with a very 
small standard deviation. Beyond $\delta_1 = 0.60$, one cannot recover the curves for $X_{3,m}^{(k)}$ given in the Figs. 3-4
 of the main article. 

$\bullet$ Second, the shape chosen for $w(\delta)$ can strongly change the resulting $X_{3,m}^{(k)}$ values. To investigate this point, we have 
fixed the two first moments of $\delta$, and chosen accordingly the parameters $\delta_1,\delta_2,\delta_{min}$ and 
$\delta_{max}$. It is found 
that $X_{3,m}^{(k)}$ can be strongly different for a gaussian weight and for a flat weight distributions. This clearly shows the strong importance of 
the interference effects, evoked in the main article, between 
the nonlinear susceptibilities of the dynamical heterogeneities corresponding to different values of $\tau$. These interference effects are strong enough to yield, 
e.g., a change in the log-log slope of $X_{3,m}^{(3)}(f/f_{\alpha} \ge 1)$ as well as a change in the values of  $X_{3,m}^{(k)}(f_{\alpha})$ by a 
factor significantly different from $1$ (i.e., larger than 2, or smaller than $1/2$). We emphasize that 
 the changes of $X_{3,m}^{(3)}(f)$ are in most cases different from those observed on $X_{3,m}^{(1)}(f)$. 
 This is the reason why it is not surprising that fitting the measured values of $X_3^{(3)}$ and of $X_3^{(1)}$ requires different 
 values of $[N_{corr}]_{av}$, as in the main article. Indeed, it is very likely that the extreme simplicity of our model cannot fully capture 
 these complicated interference effects. However, relaxing only this constraint that the values of $[N_{corr}]_{av}$ should be the same when
  fitting $X_{3}^{(3)}$ and when fitting $X_{3}^{(1)}$, we have shown, in the main 
 article, that our ADWP model is able to reproduce the salient features of the nonlinear experiments on glycerol. This is why we think that this model 
 is really relevant for showing what are the new informations about the glass 
 transition that can be drawn from nonlinear experiments in supercooled liquids. 

\begin{figure}[t]
\includegraphics*[width=9.0cm,height=6.0cm,angle=0]{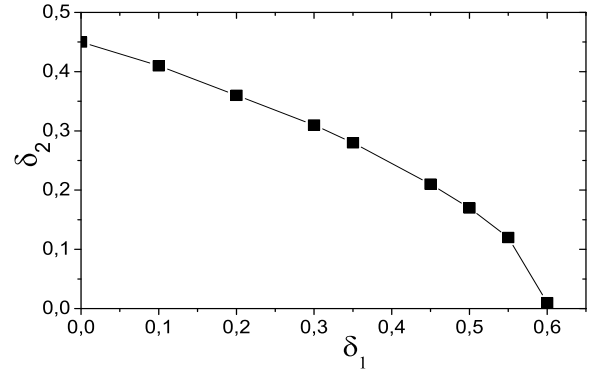}
\caption{Values of $\delta_1,\delta_2$ yielding, with a gaussian weight distribution, the same values of $X_{3,m}^{(k)}$ as those obtained in the main article with a single value of $\delta$. The line is a guide to the eyes.}
\label{Epaps-Fig1}
\end{figure}

\end{document}